\documentstyle[aps,epsfig]{revtex}
\begin{document}
%\draft

\title{Spontaneous heavy cluster emission rates using microscopic potentials}

\author{D.N. Basu\thanks{E-mail:dnb@veccal.ernet.in}}
\address{Variable  Energy  Cyclotron  Centre,  1/AF Bidhan Nagar,
Kolkata 700 064, India}
\date{\today }
\maketitle
\begin{abstract}

      The nuclear cluster radioactivities have been studied theoretically in the framework of a microscopic superasymmetric fission model (MSAFM). The nuclear interaction potentials required for binary cold fission processes are calculated by  folding in the density distribution functions of the two fragments with a realistic effective interaction. The microscopic nuclear potential thus obtained has been used to calculate the action integral within the WKB approximation. The calculated half lives of the present MSAFM calculations are found to be in good agreement over a wide range of observed experimental data.  

\end{abstract}


%\eject

      Since the first experimental observation of cluster radioactivity \cite{r1}, a lot of efforts, both experimental and theoretical, have gone into the understanding of the physics of cluster radioactivity. Lifetimes of the cluster radioactivities of radioactive nuclei have been predicted theoretically using various models and compared with existing experimental data from time to time. These models can be broadly classified as the superasymmetric fission model (SAFM) \cite{r2,r3,r4} and the preformed cluster model (PCM) \cite{r5}. In the SAFM the barrier penetrabilities are calculated assuming two asymmetric clusters. In the PCM the cluster is assumed to be formed before it penetrates the barrier and its preformation probability is also included in the calculations. Though the physics of the two approaches is apparently different, but actually they are almost  similar. Interpreting the cluster preformation probability within a fission model as the penetrability of the pre-scission part of the barrier, it was shown that the PCM is, in fact, equivalent to the fission model \cite{r6}. However, the PCM has been found to be better applicable for lighter clusters while SAFM is more apt for all cluster decays \cite{r7}.

      Both the theoretical approaches described above use either phenomenological potentials or the proximity type potentials to calculate nuclear interaction between the two fragments. The SAFM calculations using proximity type potentials or semiempirical heavy ion  potentials obtained by fitting the elastic scattering data or other phenomenological nuclear potentials for interaction between the fragments do not reproduce the observed cluster radioactivity lifetimes successfully. The SAFM using a parabolic potential approximation for the nuclear interaction potential, which is a rather unusual fragment interaction potential, however, has been found to provide reasonable estimates for the lifetimes of cluster radioactivity \cite{r4}. The PCM with various nuclear potentials have also been tried with some succes for the alpha radioactivity but was not much succesful even for a very limited number of heavier cluster decays. In the present work microscopically calculated nuclear interaction potentials have been used in the SAFM approach with reasonable success for calculating the lifetimes of cluster radioactive decays over a wide range of emitted heavy clusters from a large number of parent nuclei. The microscopic nuclear potentential obtained by double folding the cluster density distributions with realistic effective interaction is also very fundamental in nature. Moreover, the use of a single microscopic nuclear potential over a wide range of daughter and emitted cluster interaction is also aesthetically appealing. 

      In the SAFM the half life of the parent nucleus against the split into a cluster and a daughter is calculated using the WKB barrier penetration probability. The assault frequecy $\nu$ is obtained from the zero point vibration energy $E_v = (1/2)\hbar\omega = (1/2)h\nu$. The half life $T$ of the parent nucleus $(A, Z)$ against its split into a cluster $(A_e, Z_e)$ and a daughter $(A_d, Z_d)$  is given by

\begin{equation}
 T = [(h \ln2) / (2 E_v)] [1 + \exp(K)]
\label{seqn1}
\end{equation}
\noindent
where the action integral $K$ within the WKB approximation is given by

\begin{equation}
 K = (2/\hbar) \int_{R_a}^{R_b} {[2\mu (E(R) - E_v - Q)]}^{1/2} dR
\label{seqn2}
\end{equation}
\noindent
Here $\mu = mA_eA_d/A$  is the reduced mass, m is the nucleon mass, and $E(R)$ is the total interaction energy of the two fragments separated by the distance R between the centres, which is equal to the sum of nuclear interaction energy, Coulomb interaction energy and the centrifugal barrier.  The amount of energy released in the process is $Q$ and $R_a$ and $R_b$ are the two turning points of the WKB action integral determined from the equations

\begin{equation}
 E(R_a) = E(R_b) = Q + E_v
\label{seqn3}
\end{equation} 
\noindent
Energetics allow spontaneous emission of cluster only if the released energy 

\begin{equation}
 Q = M - ( M_e + M_d)
\label{seqn4}
\end{equation}
\noindent
is a positive quantity, where $M$, $M_e$ and $M_d$ are the atomic masses of the parent, the emitted cluster and the daughter nuclei, respectively,  expressed in the units of energy. Correctness of predictions for possible decay modes therefore rests on the accuracy of the ground state masses of nuclei while the reliability of the half life calculations requires proper zero point vibration energies and nuclear interaction energies.  

      In the present work the total interaction energy $E(R)$ has been evaluated using microscopic nuclear potential along with the Coulomb potential over the entire domain of interaction. The microscopic nuclear potentials have been obtained by double folding in the densities of the fragments with the finite range realistic M3Y effective interacion as

\begin{equation}
 V(R) = \int \int \rho_1(\vec{r_1}) \rho_2(\vec{r_2}) v[|\vec{r_2} - \vec{r_1} + \vec{R}|] d^3r_1 d^3r_2 
\label{seqn5}
\end{equation}
\noindent
The density distribution used for the clusters has been chosen to be of the spherically symmetric form given by

\begin{equation}
 \rho = \rho_0 / [ 1 + exp( (r-c) / a ) ]
\label{seqn6}
\end{equation}                                                                                                                                           \noindent     
where                        
 
\begin{equation}
 c = R ( 1 - \pi^2 a^2 / 3 R^2 ), ~~     R = 1.13 A^{1/3}  ~~   and ~~    a = 0.54 ~ fm
\label{seqn7}
\end{equation}
\noindent
and the value of $\rho_0$ is fixed by equating the volume integral of the density distribution function to the mass number of the cluster. The finite range M3Y effective interaction $v(s)$ appearing in the eqn. (5) is given by \cite{r8} 

\begin{equation}
 v(s) = 7999. \exp( - 4s) / (4s) - 2134. \exp( - 2.5s) / (2.5s)
\label{seqn8}
\end{equation}   
\noindent
For the direct part of the M3Y effective interaction the long range one-pion exchange potential is exactly equal to zero. As the cluster decays involve only very low energies, the finite range exchange interaction has not been considered because it is important only at higher energies \cite{r9}. This microscopic nuclear potential energy  is then used to calculate the total interaction energy $E(R)$ for use inside the WKB action integral. The two turning points of the action integral have been obtained by solving eqns.(3) using microscopic double folding potential given by eqn.(5) along with the Coulomb potential. Then the WKB action integral between the two turning points has been evaluated numerically  for calculating the half lives of the cluster decays. The zero point vibration energies used in the present calculations are same as that described by eqns.(5)  in reference \cite{r10}. The shell effects for every cluster radioactivity are implicitly contained in the zero point vibration energy due to its proportionality with the Q value, which is maximum when the daughter nucleus has a magic number of neutrons and protons. A normalisation factor of 0.9 for the microscopic nuclear potential has been used to obtain the optimum fit. The present calculation uses the experimental ground state masses for calculating the released energy Q. Whenever the experimental ground state masses are not available, it uses the theoretically calculated ground state masses from the latest mass table \cite{r11}.   

      It is important to mention here that in the analytical superasymmetric fission model (ASAFM) \cite{r2} calculations,   the entire interaction region is divided into two distinct zones. In the overlapping zone, where the distances of separation between the centres of the two fragments are below the touching radius, a parabolic form for the nuclear interaction potential has been used. And for distances beyond the touching radius only the Coulomb potential plus the centrifugal barrier for the separated fragments have been considered within a framework of a liquid drop model [LDM] two centre spherical parametrization. Treating the region beyond the touching radius as a nuclear force free zone and approximating the nuclear interaction potential to a parabolic form in the overlapping region yield  analytical expression for the WKB action integral \cite{r2}.  Although the overall uncertainty of this analytical superasymmetric fission model (ASAFM) was found to be small, neither the division of the interaction region into two distinct domains is justifiable nor the use of parabolic nuclear potential has much physical basis.

\begin{figure}[h]
\eject\centerline{\epsfig{file=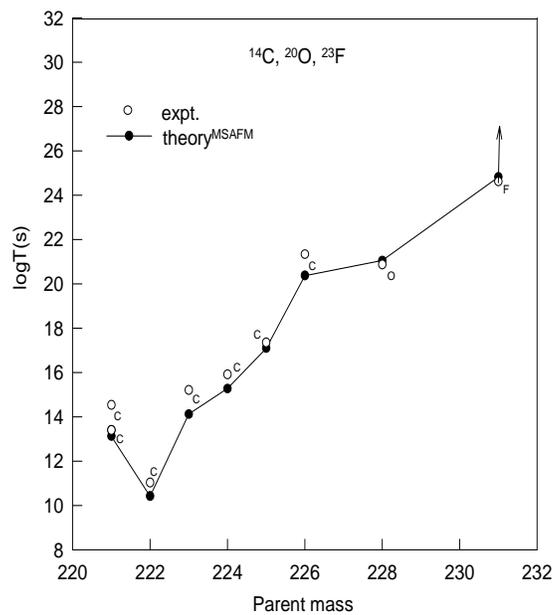,height=15cm,width=10cm}}
\caption
{Logarithmic half lives for carbon, oxygen and fluorine cluster decays plotted against parent mass number. The
continuous line connects the calculated values. The experimental data are shown by open circles, and the
arrows attached to three points indicate that these are only lower limits determined experimetally.}
\label{fig1}
\end{figure}

\begin{figure}[h]
\eject\centerline{\epsfig{file=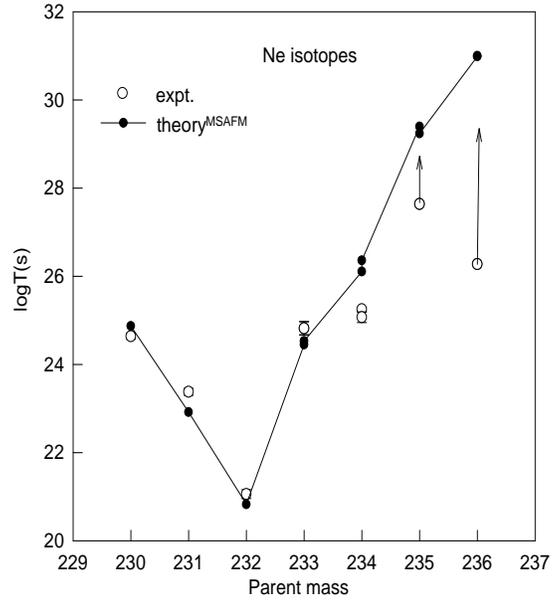,height=15cm,width=10cm}}
\caption
{Plot of logarithmic half lives for cluster decays by neon emission versus parent mass number. The continuous line connects the calculated values for different isotopes of neon. The experimental data are shown by open circles, and the arrows attached to two points indicate that these are only lower limits determined experimetally.}
\label{fig2}
\end{figure}

\begin{figure}[h]
\eject\centerline{\epsfig{file=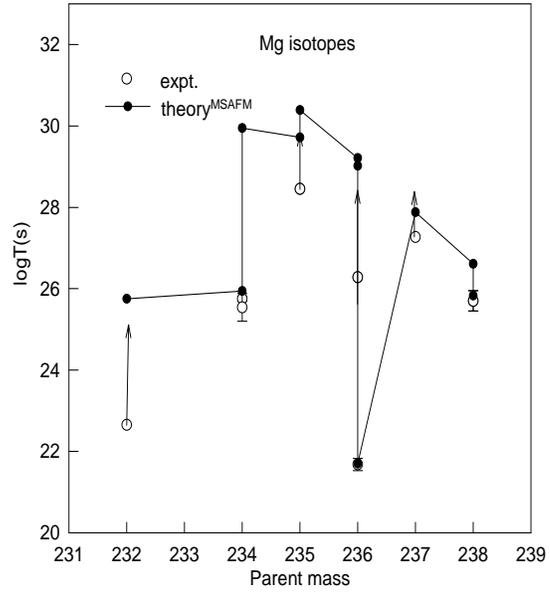,height=15cm,width=10cm}}
\caption
{Same as fig.2 but for magnesium isotopes.}
\label{fig3}
\end{figure}

\begin{figure}[h]
\eject\centerline{\epsfig{file=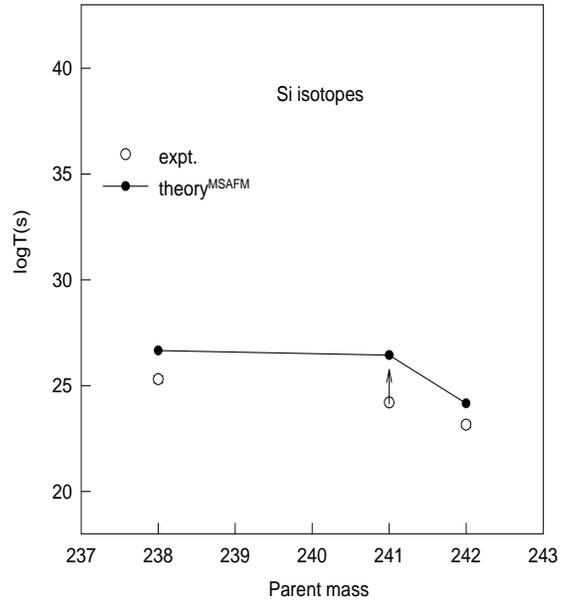,height=15cm,width=10cm}}
\caption
{Same as fig.2 but for silicon isotopes.}
\label{fig4}
\end{figure}

      In Fig.~\ref{fig1}, Fig.~\ref{fig2}, Fig.~\ref{fig3}, and  Fig.~\ref{fig4} the experimental data for logarithmic half lives \cite{r4,r12,r13,r14,r15,r16,r17,r18,r19} have been plotted against the mass numbers of parent nuclei along with the results of the present calculations for zero angular momentum of the fragments. In all the figures, the open circles depict the experimental data while the continuous line with solid circle represents the present calculations (MSAFM). The upward arrows to some experimental data points indicate that those are only the lower limits of the decay half lives determined experimentally. Fig.~\ref{fig1} contains the results of the present (MSAFM) theoretical calculations and the data points for carbon-14, oxygen-20 and fluorine-23 cluster emissions. Fig.~\ref{fig2}, Fig.~\ref{fig3}, and Fig.~\ref{fig4} represent the data and theoretical results of MSAFM calculations for cluster emissions of neon, magnesium and silicon isotopes, respectively. The decay modes and the experimental values for their half lives have been presented in Table 1. Those data that represent only the lower limits for the decay half lives have been placed at the bottom. The corresponding results of the present calculations of superasymmetric fission model with microscopic potentials (MSAFM) are also presented along with the results of ASAFM calculations of 1986 \cite{r3} and 1991 \cite{r4} so as to facilitate the comparison of the results of older calculations \cite{r4} with the present one.
 
\begin{table}
\caption{Comparison between Measured and Calculated Half-Lives}
\begin{tabular}{ccccccccccc}
Parent &       &Daughter &      &Emitted        &     & ASAFM & ASAFM &  MSAFM & Expt. &      \\ 
     &      &      &      &      &      & 1986    &  1991  &      &      &      \\ \hline
Z  & A & $Z_d$  & $A_d$  & $Z_e$ &  $A_e$ &  logT(s) & logT(s) &  logT(s) &  logT(s)  \\ \hline
87  & 221   & 81  & 207  &   6  & 14  & 15.00   & 14.37    & 13.39   & 14.52         \\
88  & 221   & 82  & 207  &   6  & 14  & 13.83   & 14.25    & 13.12   & 13.39         \\
88  & 222   & 82  & 208  &   6  & 14  & 12.56   & 11.16    & 10.41   & 11.02   \\
88  & 223   & 82  & 209  &   6  & 14  & 14.78   & 15.20    & 14.12   & 15.20   \\
88  & 224   & 82  & 210  &   6  & 14  & 17.39   & 15.95    & 15.27   & 15.90   \\
89  & 225   & 83  & 211  &   6  & 14  & 18.45   & 17.80    & 17.09   & 17.34    \\  
88  & 226   & 82  & 212  &   6  & 14  & 22.44   & 20.97    & 20.36   & 21.33   \\
90  & 228   & 82  & 208  &   8  & 20  & 22.44   & 21.95    & 21.05   & 20.86    \\  
90  & 230   & 80  & 206  & 10  & 24  & 24.86   & 25.27    & 24.87   & 24.64   \\
91  & 231   & 81  & 207  & 10  & 24  & 21.98   & 23.38    & 22.92   & 23.38   \\
92  & 232   & 82  & 208  & 10  & 24  & 20.41   & 20.81    & 20.83   & 21.06   \\
92  & 233   & 82  & 209  & 10  & 24  & 23.11   & 24.80    & 24.45   & 24.82   \\
92  & 233   & 82  & 208  & 10  & 25  & 23.44   & 25.16    & 24.53   & 24.82   \\
92  & 234   & 82  & 210  & 10  & 24  & 25.72   & 26.13    & 26.11   & 25.25   \\
92  & 234   & 82  & 208  & 10  & 26  & 26.16   & 27.05    & 26.36   & 25.07   \\
92  & 234   & 80  & 206  & 12  & 28  & 24.56   & 25.03    & 25.94   & 25.75   \\
92  & 234   & 80  & 204  & 12  & 30  & 29.15   & 29.64    & 29.95   & 25.54   \\
94  & 236   & 82  & 208  & 12  & 28  & 19.79   & 20.26    & 21.70   & 21.68   \\
94  & 238   & 82  & 210  & 12  & 28  & 24.81   & 25.29    & 26.61   & 25.70   \\
94  & 238   & 82  & 208  & 12  & 30  & 24.42   & 24.91    & 25.83   & 25.70   \\
94  & 238   & 80  & 206  & 14  & 32  & 23.69   & 24.23    & 26.66   & 25.30   \\
96  & 242   & 82  & 208  & 14  & 34  & 20.75   & 21.31    & 24.16   & 23.15   \\
91  & 231   & 82  & 208  &   9  & 23  & 24.74   & 25.89    & 24.82  &$>$24.61      \\
92  & 235   & 82  & 210  & 10  & 25  & 28.31   & 30.05    & 29.40  &$>$27.64         \\
92  & 235   & 82  & 209  & 10  & 26  & 28.40   & 30.17    & 29.24  &$>$27.64         \\
92  & 236   & 82  & 212  & 10  & 24  & 30.51   & 30.93    & 30.99  &$>$26.28         \\
92  & 236   & 82  & 210  & 10  & 26  & 30.76   & 31.65    & 31.00  &$>$26.28         \\
92  & 232   & 80  & 204  & 12  & 28  & 24.46   & 24.93    & 25.75  &$>$22.65         \\
92  & 235   & 80  & 207  & 12  & 28  & 27.33   & 29.30    & 29.72  &$>$28.45         \\
92  & 235   & 80  & 205  & 12  & 30  & 28.47   & 30.51    & 30.39  &$>$28.45         \\
92  & 236   & 80  & 208  & 12  & 28  & 27.82   & 28.29    & 29.21  &$>$26.28         \\
92  & 236   & 80  & 206  & 12  & 30  & 28.09   & 28.58    & 29.02  &$>$26.28         \\
93  & 237   & 81  & 207  & 12  & 30  & 25.84   & 27.55    & 27.88  &$>$27.27          \\
95  & 241   & 81  & 207  & 14  & 34  & 22.45   & 24.41    & 26.44  &$>$24.20           \\ 

\end{tabular} 
\end{table}

      The results of the present calculations of the MSAFM have been found to predict the general trend very well for a wide range of experimental data. The quantitave agreement with experimental data for lighter cluster emissions is excellent while that for heavier clusters is reasonable. The degree of reliability of the MSAFM predictions for cluster decay lifetimes are comparable to that of ASAFM \cite{r4}, although they are not exactly the same. It is worthwhile to mention that all the ASAFM results of 1986 and of 1991 listed in the Table 1 have been recalculated using zero-point vibration energies given by eqn.[11] of reference \cite{r3} and eqns.(5)  of reference \cite{r10} respectively. 
                                 
      The half lives for cluster-radioactivity have been analyzed with microscopic nuclear potentials which are based on profound theoretical basis. The results of the present calculations with MSAFM are in good agreement over a wide range of experimental data and are comparable to the best available theoretical calculations \cite{r4} of ASAFM which used parabolic interaction potentials that did not have any microscopic basis. Present calculations certainly put the SAFM on a firm theoretical basis. Refinements such as introduction of dissipation while tunneling through the barrier or incorporating the dynamic shape deformations in the density distributions of the clusters may further improve results. It may, however, be realised that as the first illustrative calculations using realistic microscopic cluster interaction potentials, the results of cluster radioactive decay lifetimes obtained without adjusting parameters are remarkable. In future, such calculations may therefore be extended to provide reasonable estimates of the lifetimes of nuclear decays by cluster emissions for the entire domain of exotic nuclei.

      The author is grateful to Dr. A.K. Chaudhury, Dr. K. Krishan, Dr. S. Bhattacharya and Dr. J.N. De for many helpful discussions and suggestions.

\end{document}